\documentclass[galaxies,article,accept,moreauthors,10pt,a4paper]{mdpi}
\firstpage{1}
\articlenumber{93}
\doinum{10.3390/galaxies5040093}
\pubvolume{5}
\pubyear{2017}
\copyrightyear{2017}
\history{Received: 6 October 2017; Accepted: 27 November 2017; Published: date}

\Title{Linear Polarization Properties of  Parsec-Scale AGN~Jets $^\dagger$}

\conference{{Polarised} Emission from Astrophysical Jets}

\newcommand{\ga}{\,\rlap{\raise 0.5ex\hbox{$>$}}{\lower 1.0ex\hbox{$\sim$}}\,}

\Author{Alexander B. Pushkarev $^{1,2,}$*,
        Yuri Y. Kovalev $^{2,3,4}$,
        Matthew L. Lister $^{5}$,
        Tuomas~Savolainen $^{6,7,4}$,
        Margo~F.~Aller $^{8}$,
        Hugh~D.~Aller $^8$ and
        Mary~A.~Hodge $^{5}$
        }

\AuthorNames{Alexander Pushkarev, Yuri Kovalev, Matthew Lister, Tuomas Savolainen, Margo Aller, Hugh Aller, Mary Hodge}

\address{%
$^{1}$ \quad Crimean Astrophysical Observatory, Crimea, 98409 Nauchny, Russia\\
$^{2}$ \quad Astro Space Center of Lebedev Physical Institute, Profsoyuznaya 84/32, 117997 Moscow, Russia; yyk@asc.rssi.ru\\
$^{3}$ \quad Moscow Institute of Physics and Technology, Dolgoprudny, Institutsky per., 9, 141700~Moscow~Region,~Russia\\
$^{4}$ \quad Max-Planck-Institut f\"ur Radioastronomie, Auf dem H\"ugel 69, 53121 Bonn, Germany; tsavolainen@mpifr-bonn.mpg.de\\
$^{5}$ \quad Department of Physics and Astronomy, Purdue University, 525 Northwestern Avenue, West~Lafayette,~IN~47907,~USA; mlister@purdue.edu (M.L.L.); hodge2@perdue.edu (M.A.H.)\\
$^{6}$ \quad Mets\"ahovi Radio Observatory, Aalto University, Mets\"ahovintie 114, FI-02540 Kylm\"al\"a, Finland;\\
$^{7}$ \quad Department of Electronics and Nanoengineering, Aalto University, PL 15500, FI-00076 Aalto, Finland;\\  
$^{8}$ \quad Department of Astronomy, University of Michigan, 311 West Hall, 1085 S. University Avenue, Ann~Arbor,~MI~48109,~USA; mfa@umich.edu (M.F.A.);  haller@umich.edu (H.D.A.)\\
}

\corres{Correspondence: pushkarev.alexander@gmail.com}

\abstract{
We used 15 GHz multi-epoch Very Long Baseline Array (VLBA) polarization sensitive 
observations of 484 sources within a time interval 1996--2016 from the MOJAVE 
program, and also from the NRAO data archive. We have analyzed the linear polarization 
characteristics of the compact core features and regions downstream, and their changes 
along and across the parsec-scale active galactic nuclei (AGN) jets. We detected a 
significant increase of fractional polarization with distance from the radio core 
along the jet as well as towards the jet edges. Compared to quasars, BL Lacs have a 
higher degree of polarization and exhibit more stable electric vector position angles 
(EVPAs) in their core features and a better alignment of the EVPAs with the local jet 
direction. The latter is accompanied by a higher degree of linear polarization, 
suggesting that compact bright jet features might be strong transverse shocks, which 
enhance magnetic field regularity by compression.
}


\begin{document}
\section{Introduction}

Despite substantial progress in the understanding of the phenomenon of the jets of
active galactic nuclei (AGN) achieved over the last several decades with the method
of very long baseline interferometry (which probes jets from sub- to several hundred
parsec scales) the question of how jets are launched, confined and collimated remains
an active area of research. The dominant acceleration mechanisms, which could be
steady or impulsive in nature, are still unknown. All these questions are expected to
be tightly connected with the key agent of jet dynamics, the magnetic (B) field.
Construction of the Very Long Baseline Array (VLBA) in 1994 has allowed polarimetric
observations of large samples of AGN jets on a regular basis facilitating detailed
studies of the polarization characteristics of jet synchrotron radio emission.

VLBA polarization maps provide crucial information on the configuration of the magnetic
field associated with an outflow, its regularity and orientation with respect to the
local jet direction. In this publication, we analyze the polarization properties of a
large sample of AGN jets mainly observed within the MOJAVE program \citep{Lister09}
with the VLBA at 15~GHz.

\section{Observational Data and the Sample}

The data consist of 5410 polarization sensitive VLBA observations of 484 AGNs
at 338~individual epochs between 1996 January 19 and 2016 December 26. The sources are drawn from
a number of samples: the complete flux density-limited MOJAVE 1.5 sample \cite{Lister15},
the joint gamma-ray and radio-selected sample 1FM \cite{Lister11}, the VLBA 2~cm survey
\cite{Kellermann98}, the MOJAVE low-luminosity sample \cite{Lister13}, the 3-rd EGRET 
gamma-ray catalog \cite{Hartman99}, and the 3FGL {\it Fermi} LAT gamma-ray catalog 
\cite{Acero15}. Most of the sources (80\%) have been detected at high energies by the
{\it Fermi} LAT instrument \cite{Acero15}. All targets were bright enough ($\ga$50~mJy)
at 15~GHz for direct fringe detection on short integration times. We reduced the data
in AIPS software package using standard techniques and performed imaging in Difmap
\cite{Shepherd97}.

The overwhelming majority (88\%) of the VLBA observations were done within the MOJAVE
program \cite{Lister09}, while the rest were obtained from the NRAO archive to increase
the number of epochs for certain sources. The source cadence in the MOJAVE program is
individually determined by the proper motion of jet knots and varies from about a month
to two years. The median number of VLBA observation epochs per source is seven, although
54 sources have more than 20 epochs, with a maximum of 133 epochs for BL Lac. Since the
beginning of the MOJAVE program in 2002, the noise level of the Stokes Q and U maps
improved by a factor of $\sim$3 and currently reaches a typical value of
$\sim$0.1~mJy~beam$^{-1}$, with a bit rate of 2~Gbps and 2-bit sampling. The
corresponding linear polarization and total intensity maps with a more detailed
description of image characteristics are available in \cite{Lister17}.

The sample is strongly dominated by flat-spectrum radio quasars (71\%), with a
significant fraction of BL Lacertae objects (20\%), and a small fraction of radio galaxies
(7\%). The rest of the sample (2\%) is comprised of optically unidentified sources. The
redshifts are currently known for 443 objects (91\%), ranging from 0.00436 for the galaxy
M87 to 3.636 for the BL Lac object 1549+089 and corresponding to scale factors ranging
from 0.01 to 7.35~pc~mas$^{-1}$, respectively. Taking into account that a typical angular
resolution of the VLBA observations at 15~GHz is of the order of 1~mas, path lengths along
the constructed ridgelines range from 1 to 57 mas and viewing angles are of the order of
few degrees for blazars and may reach up to few tens of degrees for radio galaxies
\cite{Pushkarev17}. This implies that we probe absolute linear scales from sub- to
hectoparsecs of the collimated AGN outflows.

\section{Results}

For the purposes of our analysis we produced maps of polarized intensity
$P=(Q^2+U^2)^{1/2}$, electric vector position angle $\chi=0.5\,\textrm{atan}(U/Q)$, and
fractional polarization $m=P/I$, where $I$ is the total intensity. Since the noise in a
$P$ image is non-uniform and follows a Ricean distribution, we~adopted a conservative
approach of estimating the detection limit based on a level at which spurious $P$ signals
appear in blank sky region of each map. Typically, this corresponds roughly to a
$3\sigma$ level, but~in about 16\% of the observations it exceeds 5 times $\sigma$ due
to a high peak in a total intensity map or high instrumental polarization.

\begin{figure}[H]
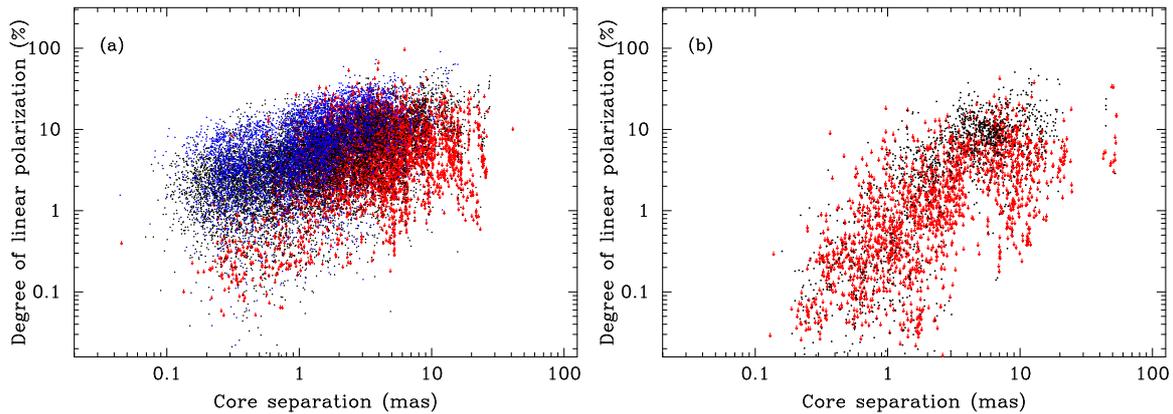

\centering
\includegraphics[width=5.4 cm,angle=-90]{m_vs_r_BQ.eps}\hspace{1mm}
\includegraphics[width=5.4 cm,angle=-90]{m_vs_r_G.eps}
\caption{(\textbf{a}) Fractional polarization at positions of components in total
         intensity with distance from the 15~GHz core for quasars (black) and BL
         Lacertae objects (blue). Dots show measurements, while red arrows represent upper limits.
         (\textbf{b}) Same for radio galaxies.}
\label{f:m_vs_r}
\end{figure}

We extracted fractional linear polarization values from the constructed $m$-maps at the
positions of total intensity components that were derived from structure model fitting
performed in Difmap. If the polarization intensity of a component was lower than the
noise level, we calculated a corresponding upper limit. In Figure~\ref{f:m_vs_r} we
plot the degree of polarization of the jet components of all sources separated by optical
classification at all available epochs as a function of distance to the VLBA core. It
shows a tendency for fractional polarization to increase with core separation implying
that the magnetic field becomes more regular down the jet. Polarization of radio galaxies
is weaker, especially for jet components within a few milliarcseconds from the core, with
a significantly larger fraction of upper limits (53\%) compared to quasars (23\%) and
BL Lacs (22\%).

\begin{figure}[H]
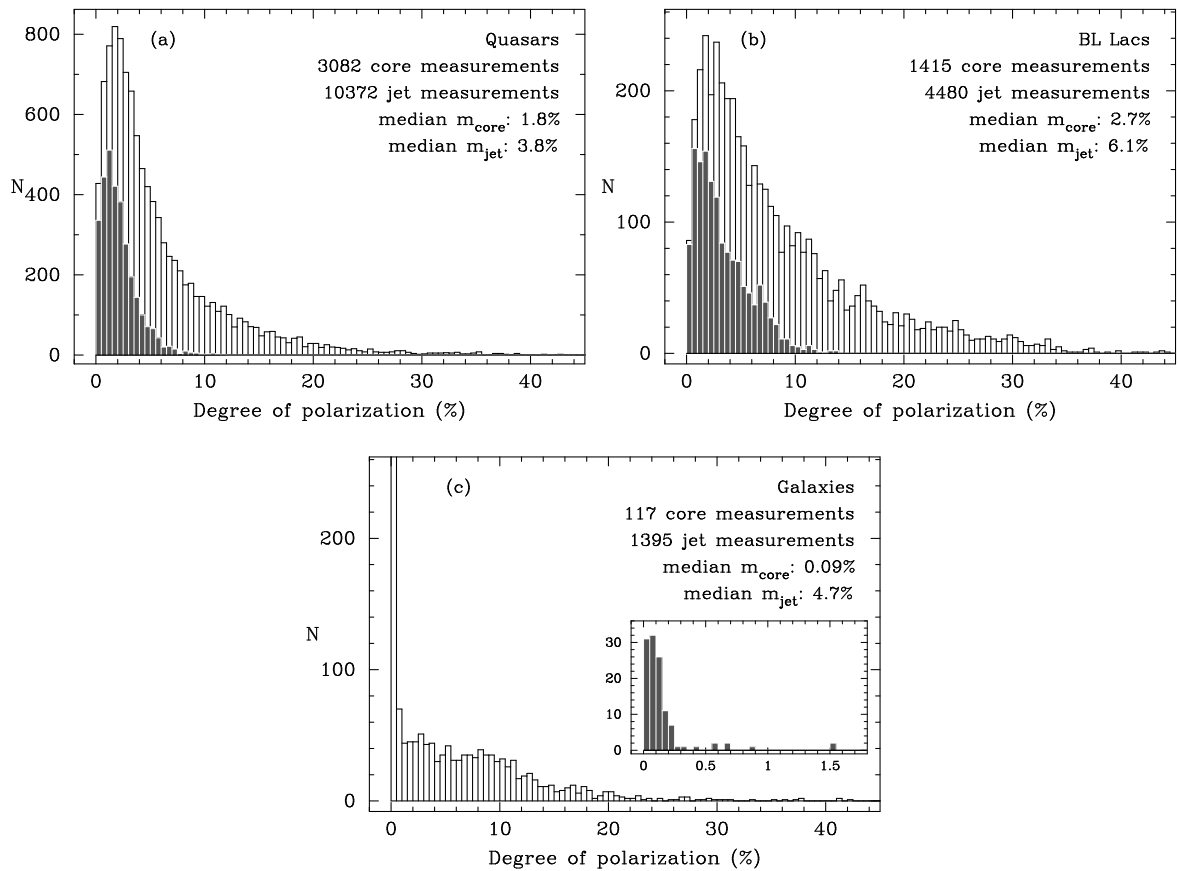

\centering
\includegraphics[width=5.5 cm,angle=-90]{hist_m_core_jet_Q.eps}\hspace{1mm}
\includegraphics[width=5.5 cm,angle=-90]{hist_m_core_jet_B.eps}\vspace{4mm}
\includegraphics[width=5.5 cm,angle=-90]{hist_m_core_jet_G.eps}
\caption{(\textbf{a}) Histograms of fractional polarization of core (filled
         gray bins) and jet components (empty bins) for quasars, (\textbf{b})
         for BL Lacs, (\textbf{c}) for radio galaxies.}
\label{f:m_hist}
\end{figure}

Core features typically have the highest levels of polarized flux density in the jet
but lower fractional linear polarization. The cores, being partially optically thick,
have typical degree of polarization on levels of a few per cent in quasars and BL Lacs,
while the cores of radio galaxies are weakly polarized, with $m_\textrm{core}<0.5$\%
(Figure~\ref{f:m_hist}). Low- and intermediate SED peak BL Lacs have the most highly
polarized cores. In only 30 out of 5410 total epochs $m_\textrm{core}>10$\%. Both core
and jet components of BL Lacs, on average, are more polarized than those of quasars.

\begin{figure}[H]
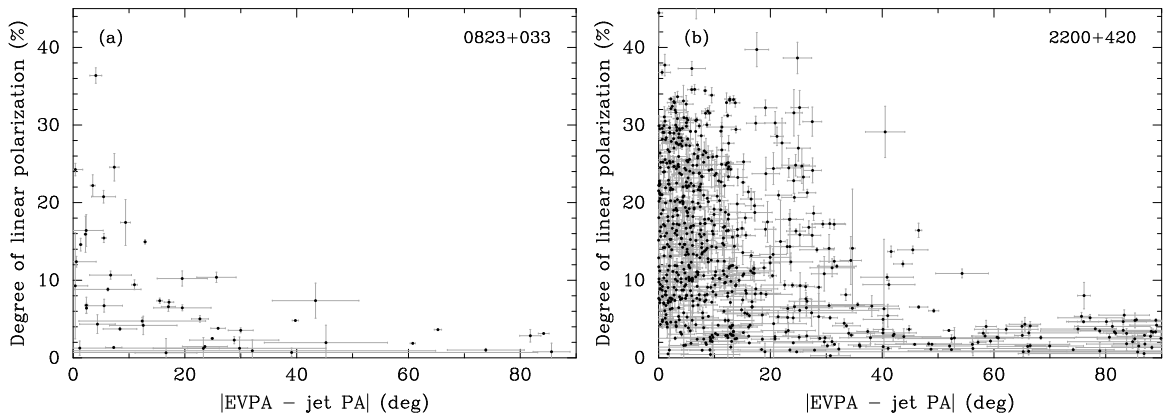

\centering
\includegraphics[width=5.4 cm,angle=-90]{0823+033_evpa_jetpa_m.eps}\hspace{1mm}
\includegraphics[width=5.4 cm,angle=-90]{2200+420_evpa_jetpa_m.eps}
\caption{(\textbf{a}) Negative correlation between degree of polarization
         of total intensity components and absolute deviation of the electric vector position angles (EVPA) from the
         local jet direction for BL Lac object 0823 + 033. (\textbf{b}) Same for BL~Lac.}
\label{f:evpa_jetpa_m}
\end{figure}

We have also found that roughly 40\% of the AGN jet cores show a tendency for a preferred
EVPA direction over time. BL Lac cores have more stable EVPAs than those of quasars and
show a tendency to be aligned with the inner parsec-scale jet. Similar behaviour is
detected for the EVPA of jet components of BL Lacertae objects, which tend to be aligned
with the local jet direction. In~contrast, quasars and radio galaxies do not show such a
trend. This confirms the findings of earlier studies, e.g.,~\cite{Gabuzda00,Lister05}. In~many
BL Lacs, high degrees of polarization are associated with low deviations of EVPAs from
the jet direction (Figure~\ref{f:evpa_jetpa_m}).

\begin{figure}[H]
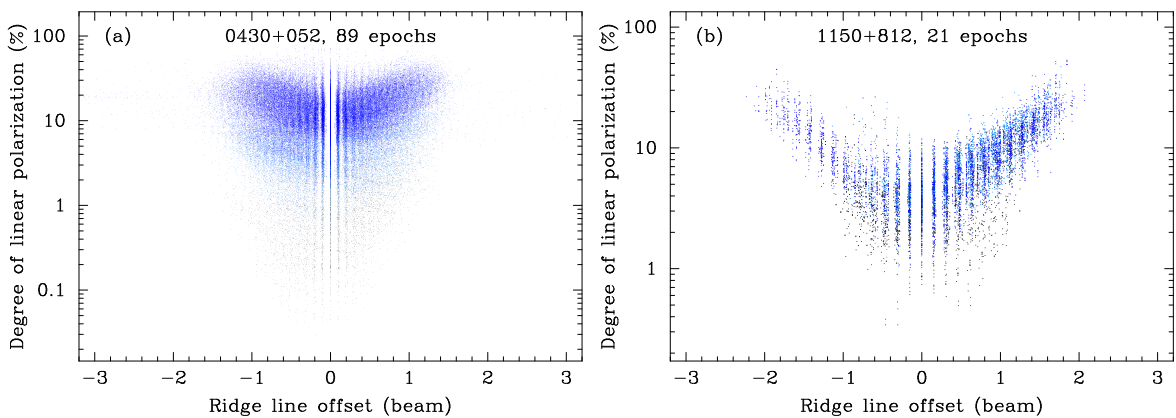

\centering
\includegraphics[width=5.4 cm,angle=-90]{0430+052.u.all_epochs_m_log_cut.eps}\hspace{1mm}
\includegraphics[width=5.4 cm,angle=-90]{1150+812.u.all_epochs_m_log_cut.eps}
\caption{(\textbf{a}) Fractional polarization from slices transverse to the local jet
         direction as a function of offset from total intensity ridgeline for the
         radio galaxy 3C~120 (0430+052). Black, cyan and blue dots show measurements
         at distances $r<1$~mas, $1<r<3$~mas, and $r>3$~mas from the 15~GHz core
         measured along ridgeline, respectively. (\textbf{b}) Same for the quasar
         1151+812.}
\label{f:m_cuts}  
\end{figure}

Fractional polarization maps taken within our program, and also in individual source studies,
e.g., \cite{Pushkarev05}, often manifest clear increase of $m$-values towards the jet edges.
To analyze changes in degree of polarization across the jet, we constructed jet ridgelines
in total intensity following the procedure described in \cite{Pushkarev17}, and made
$m$-slices transverse to the local jet direction. In Figure~\ref{f:m_cuts} we present
examples of two sources, the radio galaxy 0430 + 052 and quasar 1150 + 812, that reveal rich
polarization structure at 15~GHz and have many epochs of observations. The sources show a
characteristic V-shaped profile in the $m$-cuts (where the $x$-axis is a measure of offset
from the ridgeline of the jet in units of restoring beams). Closer to the jet edges, the
degree of polarization increases up to a few tens of per cent. Many other sources show
similar tendency suggesting that this effect is quite~common.

\section{Discussion}

Since a jet expands with distance, magnetic field decreases as $B\propto r^{-b}$
($b=1$ for a pure toroidal field, $b=2$ for a pure axial field) but becomes more regular,
as is evident from the observed increase in fractional polarization. One possible explanation
of this trend is that magnetohydrodinamic turbulence is expected to decrease with distance
from relativistic shocks and also down the jet~\cite{Bottcher16}. The observed increase of
the degree of polarization towards the jet edges is likely due to the superposition of
synchrotron emission from regions with different magnetic field orientations resulting in
depolarization, which is more efficient closer to the jet axis and weaker to the edges,
where the emission layer is thinner.

The VLBA core features have a degree of polarization within 10\% implying that synchrotron
radiation from these regions is optically thick, while it is optically thin in the jet as
$m_\textrm{jet}$ often exceeds this limit. In BL Lacs, the observed increase in alignment
of EVPA with jet direction (perpendicular B-field) accompanied by a higher fractional
polarization suggests that the bright jet components might be shock fronts that enhance
magnetic field orderliness by compression. Alternatively, it may be interpreted in terms
of a toroidal component of a large-scale helical B-field associated with the jet, though
this scenario faces difficulties explaining inverse dependence between
$|\textrm{EVPA}-\textrm{jet PA}|$ and $m$ detected in BL Lacs.

\vspace{6pt}

\acknowledgments{The MOJAVE project was supported by NASA-Fermi grants
NNX08AV67G, NNX12A087G, and NNX15AU76G. MFA was supported in part by
NASA-Fermi GI grants NNX09AU16G, NNX10AP16G, NNX11AO13G, NNX13AP18G
and NSF grant AST-0607523. This research has made use of data from
the MOJAVE database that is maintained by the MOJAVE team
\cite{Lister09}. This work made use of the Swinburne University
of Technology software correlator \cite{Deller11}, developed as
part of the Australian Major National Research Facilities Programme
and operated under licence. YYK and ABP are partly supported by the
Russian Foundation for Basic Research (project 17-02-00197), the
government of the Russian Federation (agreement 05.Y09.21.0018), and
the Alexander von Humboldt Foundation. TS was supported by the Academy
of Finland projects 274477 and 284495. The National Radio Astronomy 
Observatory is a facility of the National Science Foundation operated 
under cooperative agreement by Associated Universities, Inc.}

\authorcontributions{M.L.L., Y.Y.K., A.B.P. and T.S. reduced and analyzed the data.
M.F.A. and H.D.A. obtained simultaneous single dish polarization observations for
calibration of the VLBA measurements. M.A.H. calibrated EVPAs for a number of experiments.}

\conflictsofinterest{The authors declare no conflict of interest.}

\abbreviations{The following abbreviations are used in this manuscript:\\

\noindent
\begin{tabular}{@{}ll}
AGN    & Active Galactic Nucleus\\
AIPS   & Astronomical Image Processing System\\
EVPA   & Electric Vector Position Angle\\
MOJAVE & Monitoring Of Jets in Active galactic nuclei with VLBA Experiments\\
NRAO   & National Radio Astronomy Observatory\\
SED    & Spectral Energy Distribution\\
VLBA   & Very Long Baseline Array
\end{tabular}}

\reftitle{References}

\end{document}